\documentclass[onecolumn,showpacs,showkeys,preprintnumbers,
amsmath,amssymb,aps,floatfix,preprint]{revtex4}

\usepackage{graphicx}
\usepackage{multirow}
\usepackage{times}
\usepackage{amssymb}
\usepackage{dcolumn}%
\usepackage{bm}%
\usepackage{hyperref}%

\hypersetup
{
	colorlinks=true,
	linkcolor=blue,
	citecolor=blue
}

\begin{document}

\title[]{Competition of  the connectivity with the local and the global order in polymer melts and crystals}
\author{S.\@ Bernini}
\affiliation{Dipartimento di Fisica ``Enrico Fermi'', 
Universit\`a di Pisa, Largo B.\@Pontecorvo 3, I-56127 Pisa, Italy}
\author{F.\@ Puosi\footnote{Present 
 address: Laboratoire Interdisciplinaire de Physique, Universit\`e Joseph Fourier Grenoble, CNRS,	
 Ê38402 Saint Martin d'H\`eres, France.}}
\affiliation{Dipartimento di Fisica ``Enrico Fermi'', 
Universit\`a di Pisa, Largo B.\@Pontecorvo 3, I-56127 Pisa, Italy}
\author{M.\@ Barucco}
\affiliation{Dipartimento di Fisica ``Enrico Fermi'', 
Universit\`a di Pisa, Largo B.\@Pontecorvo 3, I-56127 Pisa, Italy}
\author{D.\@ Leporini}
\email{dino.leporini@df.unipi.it}
\affiliation{Dipartimento di Fisica ``Enrico Fermi'', 
Universit\`a di Pisa, Largo B.\@Pontecorvo 3, I-56127 Pisa, Italy}

\date{\today}

\begin{abstract}
The competition between  the connectivity and the local or global order in model fully-flexible chain molecules is investigated by molecular-dynamics simulations.  States  with both missing  (melts) and high (crystal) global order are considered.  Local order is characterized within the first coordination shell (FCS) of a tagged monomer and found to be lower than in atomic systems in both melt and crystal. The role played 
by the bonds linking the tagged monomer to FCS monomers  (radial bonds), and the bonds linking two FCS monomers (shell bonds) is investigated. The detailed analysis in terms of  Steinhardt's orientation order parameters $Q_l$ ($l=2-10$)  reveals that increasing the number of shell bonds decreases the FCS order in both melt and crystal. Differently, the FCS arrangements organize the radial bonds.  Even if the molecular chains are fully flexible, the distribution of the angle formed by adjacent radial bonds exhibits sharp contributions at  the characteristic angles $\theta \approx 70^\circ, 122^\circ, 180^\circ$. The fractions of adjacent radial bonds with $\theta \approx  122^\circ, 180^\circ$ are enhanced by the global order of the crystal, {whereas the fraction with $ 70^\circ \lesssim \theta \lesssim 110^\circ$  is nearly unaffected by the crystallization}. Kink defects, i.e. large lateral displacements of the chains, are evidenced in the crystalline state.
\end{abstract}

\maketitle

\section{Introduction}
Understanding the progressive solidification of  systems -such as polymers, colloids, metallic glasses, and liquids-
to get to the amorphous glassy state by avoiding the possible crystallization is a major scientific challenge \cite{EdigerHarrowell12,AngelNgai00,DebeStilli2001,Richert02,BerthierBiroliRMP11}. The huge slowing down of the dynamics, the non-exponential relaxation and the broad distribution of relaxation times, the spatial distribution of mobility leading to dynamic heterogeneity are distinctive phenomena of the glass transition which are lively debated. 
A crucial aspect of the solidification leading to a glass is that it is associated only to subtle structure changes. This led to develop theories disregarding the microscopic organization and interpreting the structural arrest in terms of an order-disorder dynamical phase transition between active fluid states and inactive states where structural relaxation may be completely arrested \cite{GarrahanChandlerScience09}.
A different line of thought suggests that structural aspects matter in the dynamical behaviour of glassforming systems. This includes the Adam-Gibbs derivation of the structural relaxation \cite{AdamGibbs65} - built on the thermodynamic notion of the configurational entropy \cite{GibbsDiMarzio58} - and developments reviewed in ref.\cite{DudowiczEtAl08}, the mode- coupling theory \cite{GotzeBook} and extensions \cite{SchweizerAnnRev10}, the random first-order transition theory (RFOT) \cite{WolynesRFOT07}, the frustration-based approach \cite{TarjusJPCM05}, as well as the so-called  elastic models  \cite{Dyre06,Puosi12}.  The search of a link between structural ordering and slow dynamics motivated several studies in liquids \cite{NapolitanoNatCom12,EdigerDePabloNatMat13,RoyallPRL12}, colloids \cite{TanakaNatMater08,TanakaNatCom12} and polymeric systems \cite{DePabloJCP05,GlotzerPRE07,LasoJCP09,BaschnagelEPJE11,MakotoMM11}.

While global order is virtually absent in macroscopically-disordered systems like glasses and liquids, {\it local} order is present in both disordered and ordered phases \cite{FrenkelPRL95} with  differences well-known  in {\it atomic} systems \cite{TorquatoStilliPRE02,AsteJPCM05,AstePRE05}. Here, we report on the local order in polymers and oligomers where the chain connectivity creates constraints and then is expected to compete with ordering phenomena. States  with both missing  (melts) and high (crystal) global order are considered and compared. 
Our study is motivated by the fact that  the differences in local order between {\it atomic} systems and {\it connected} systems are still not well characterized. It is known by a molecular-dynamics (MD) study of freely jointed chains of tangent hard spheres (HS) that, as in atomic systems \cite{AsteJPCM05},  there is no evidence of hexagonal close packed and face- centered cubic local order \cite{LasoJCP09}. Signs of icosahedral order have been revealed in a model polymeric system \cite{DePabloJCP05} and polymer-tethered nanospheres \cite{GlotzerPRE07}. Locally bundled bonds exhibiting orientational order have been also reported \cite{MakotoMM11}.

\begin{figure}[t]
\begin{center}
\includegraphics[width=0.5\linewidth]{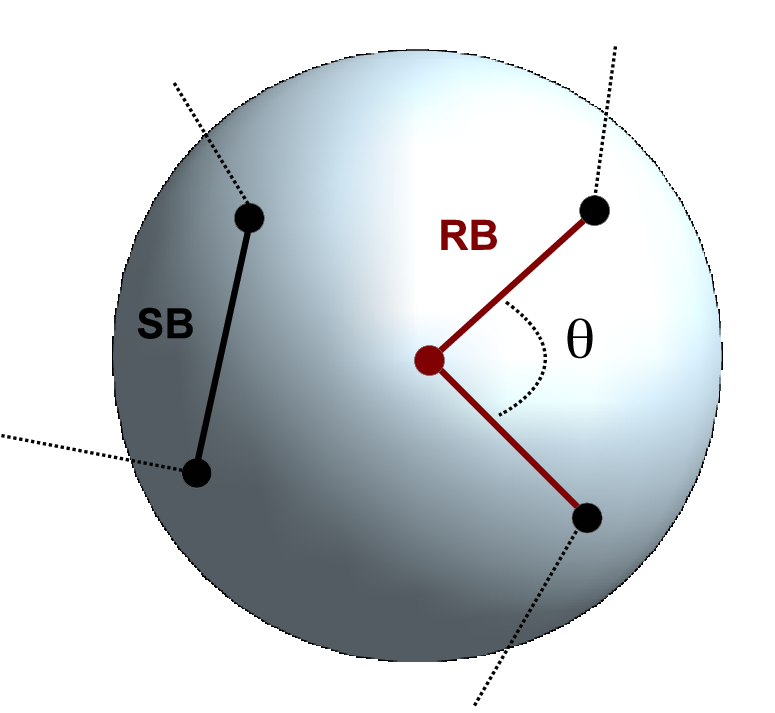} \\
\end{center}
\caption{Schematic view of the region of interest to assess the local order. The region, denoted by FCSR, includes a tagged monomer (red dot), and its first coordination shell (FCS), sketched as a spherical surface. The actual monomer diameter, roughly the FCS radius, is comparable to the bond length in our model. Depending if the tagged monomer is a chain-end or an inner monomer (as pictured), it is linked by one or two radial bonds (RB, red lines), respectively, to monomers located in FCS (black dots). Bonds linking two FCS monomers are referred to as shell bonds (SB, black lines). Dotted lines are possible additional bonds.}
 \label{fig0}
\end{figure}

The paper presents a thorough MD study of model polymer and oligomer melts with fully-flexible linear chains. Both the instantaneous and the inherent dynamics, localizing the system in mechanically-equilibrated configurations deprived of thermal vibrations, are considered \cite{DebeStilli2001,StillingerScience95}. 
To assess the local order, we consider the volume bounded by the first coordination shell (FCS) of a tagged monomer, i.e. the region including FCS and the tagged monomer (see Fig.\ref{fig0}).  From now on, this region of interest will be denoted by FCSR.
Two kind of bonds are present in FCSR (see Fig.\ref{fig0}): i)  the bonds linking the tagged monomer to FCS monomers, henceforth to be referred to as radial bonds (RB), and ii) the bonds linking each other two FCS monomers, henceforth to be referred to as shell bonds (SB). One or two RBs are present per FCSR, depending if the tagged monomer is located in either a chain-end or the inner part of the chain, respectively. More SBs than RBs per FCSR are expected and, in fact,  up to seven SBs per FCSR are found (see below). On this basis, different strengths in the SBs and RBs competition with the FCS ordering are expected. 
The FCS order is investigated by Steinhardt's  order parameters \cite{Steinhardt83}, a measure of the orientational order from global to local scales \cite{Steinhardt83,FrenkelPRL95,AsteJPCM05,VoronoiMeckeJCP13,ReviewOrderParameter,TorquatoStilliPRE02}.  

The paper is organized as follows. In Sec. \ref{numerical} the polymer model and the MD algorithms are presented. The results are discussed in Sec.\ref{resultsdiscussion}. Finally, the main conclusions are summarized in Sec.\ref{conclusions}.

\section{Methods}
\label{numerical}

A coarse-grained model of a melt of linear fully-flexible unentangled polymer chains with fixed bond lenght and $M$ monomers each is used. Oligomers ($M=3$) and short polymers ($M=10$) are considered.  The system has $N=501$ ($M=3$) or $N=500$ ($M=10$) monomers. All the quantities evaluated by the instantaneous configurations, apart from $Q_{l,global}$ (see Eq.\ref{Ql_global}), are found to be unaffected by increasing the number of monomers up to $N=2000$. Nonbonded monomers at a distance $r$ belonging to the same or different chains interact via a truncated Lennard-Jones (LJ) potential:
\begin{equation}
U_{LJ}(r)=\epsilon\left [ \left (\frac{\sigma^*}{r}\right)^{12 } - 2\left (\frac{\sigma^*}{r}\right)^6 \right]+U_{cut}
\end{equation}
where $\sigma^*=2^{1/6}\sigma$ is the position of the potential minimum with depth $\epsilon$, and the value of the constant $U_{cut}$ is chosen to ensure $U_{LJ}(r)=0$ at $r \geq r_c=2.5\,\sigma$. Bonded monomers are constrained to a distance $b=0.97\,\sigma$ by using the RATTLE algorithm \cite{allentildesley}. 
All quantities are in reduced units: lenght in units of $\sigma$, temperature in units of $\epsilon/k_B$ and time in units of $\sigma\sqrt{\mu/\epsilon}$ where $\mu$ is the monomer mass. We set $\mu=k_B=1$. 
The range of the investigated temperatures of the melt is $0.5 \le T \le 1$,  i.e. above the MCT critical temperature $T_c \simeq 0.45$ \cite{sim} ($T_c \sim 1.15 \,T_g$). The number density of the monomers is $\rho=0.984$. The crystalline state of the decamer ($M=10$) has temperature $T=0.75$ and density $\rho = 1.086$.
$NPT$ and $NTV$ ensembles have been used for equilibration runs, while $NVE$ ensemble has been used for production runs for a given state point
(NVT: constant number of particles, volume and temperature; NVE: constant number of particles, volume and energy; NPT constant number of particles, pressure and temperature). $NPT$ and $NTV$ ensembles are studied by the extended system method introduced by Andersen \cite{Andersen80} and Nos\'e \cite{NTVnose}. 
The numerical integration of the augmented Hamiltonian is performed through the multiple time steps algorithm, reversible Reference System Propagator Algotithm (r-RESPA) \cite{respa}. For further details, see e.g. ref.\cite{OurNatPhys}.

At non-zero temperature monomers vibrate around their equilibrium positions; such fast movements make it  difficult to characterize the arrangement of monomers. In order to remove vibrations  one resorts to the so-called inherent structures (IS) by mapping the configurations of the simulated trajectory into the corresponding local minimum of the potential energy  \cite{StillingerScience95}. The conjugate-gradient method is used to minimize the configurational energy as a function of the $3N$ particles coordinates \cite{NumRecC}. Henceforth, a physical quantity $X$ will be denoted as $X^{IS}$ if evaluated in terms of IS configurations.

\begin{figure}[t]
\begin{center}
\includegraphics[width=0.5\linewidth]{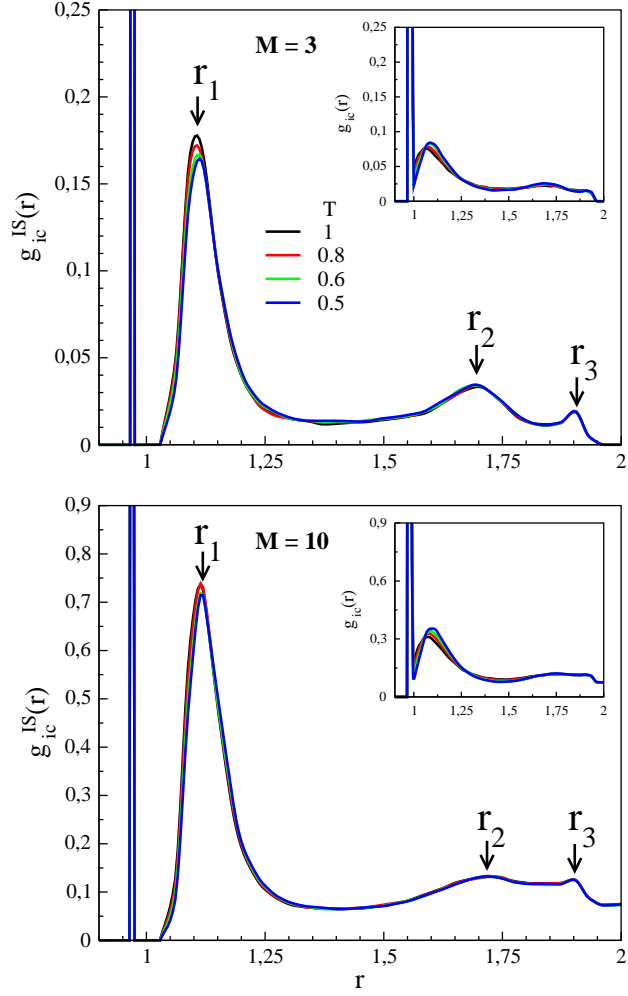} \\
\end{center}
\caption{Intrachain part of the inherent radial distribution function $g_{ic}^{IS}(r)$ for trimers (top) and decamers (bottom) at the indicated temperatures. Inset: corresponding instantaneous distribution $g_{ic}(r)$.  The delta-like peak is due to the bonded monomer at $r=b=0.97$. The maximum occurs at $r = r_1$ with $r_1$ close to the equilibrium distance of two non-bonded monomers $\sigma^*=2^{1/6} \simeq 1.12$. Note that the distribution vanishes at $r \sim 2$ for trimers but it extends farther for decamers.}
 \label{fig1}
\end{figure}
\begin{figure}[t]
\begin{center}
\includegraphics[width=0.5\linewidth]{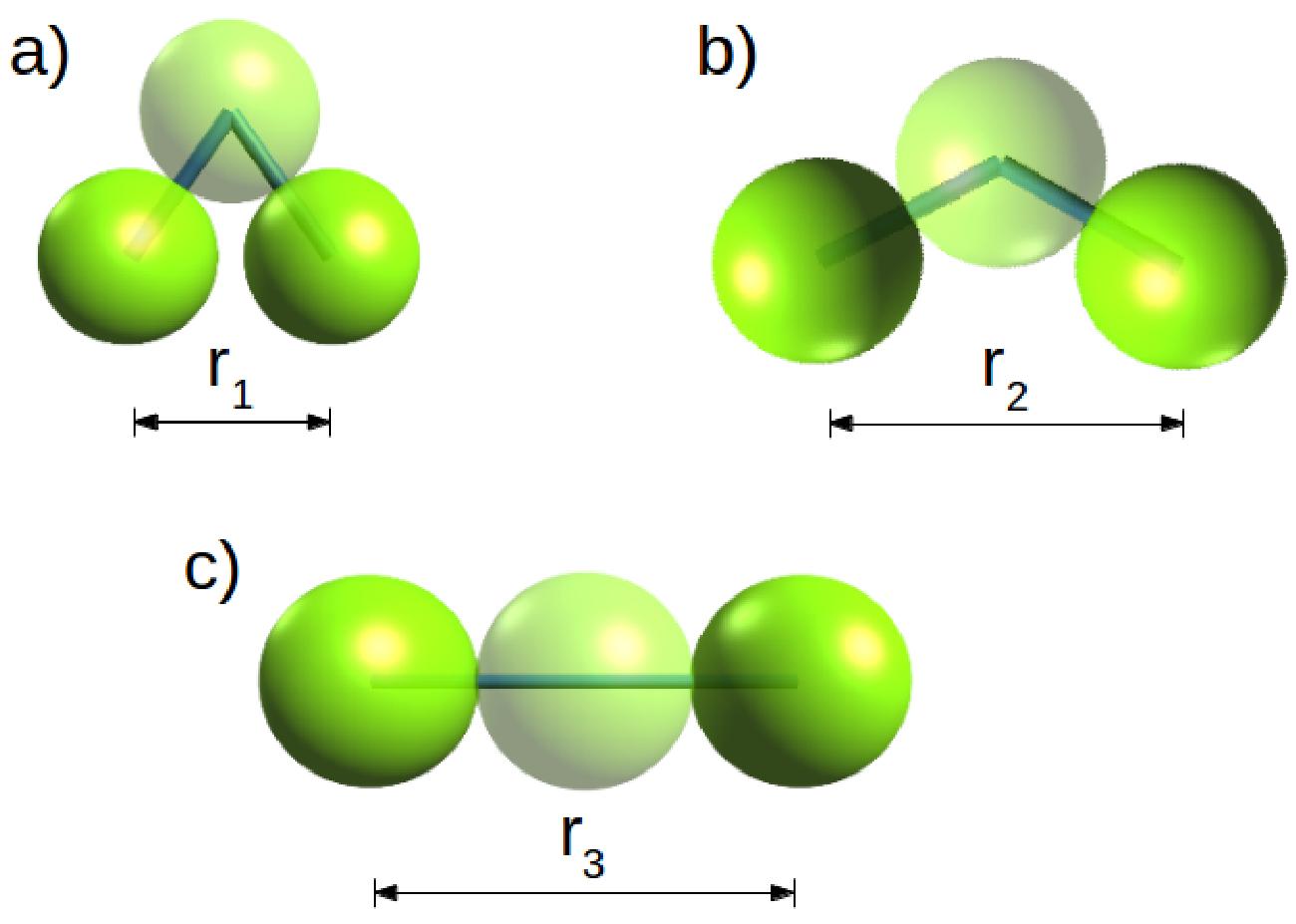} \\
\end{center}
\caption{Monomer arrangements corresponding to the three peaks of the intrachain part of the radial distribution function in Fig.\ref{fig1}. For clarity reasons,  the chain fragments with  the relevant monomers and bonds are only plotted. The bonds in each fragment are the RBs of the central monomer (see Fig.\ref{fig0}).
$r$ and $\theta$ denote the distance between the two non-bonded monomers and the bond-bond angle, respectively. a) folded conformation ($r_1 \approx \sigma^* = 1.12$, $\theta_1\approx 70^\circ$); b) partially folded conformation ($r_2 \approx 1.69$, $\theta_2 \approx 122^\circ$);  c) linear conformation ($r_3 \approx 1.94$, $\theta_3\approx 180^\circ$). }
 \label{figConfoBond}
\end{figure}

\section{Results and discussion}
\label{resultsdiscussion}

Sections  \ref{radial}, \ref{bondbond} and \ref{orderpar} present and discuss the results concerning the melt state.
Sec. \ref{radial} illustrates  the intrachain radial distribution function, evidencing characteristic arrangements of monomers linked by adjacent bonds of the chains, i.e. RBs (see Fig. \ref{fig0}). The related RB angular ordering is presented and discussed in Sec. \ref{bondbond}. The previous results are compared with the ones drawn by the analysis of the FCS orientational order in terms of the Steinhard's  order parameters in Sec.\ref{orderpar} where the role of the SBs is also analysed. 

Finally, in Sec.\ref{global} the results concerning the melt and the crystalline states are compared.

\subsection{Intrachain radial distribution function}
\label{radial}
Fig. \ref{fig1} shows the intrachain part of the radial distribution function $g_{ic}^{IS}(r)$ for trimers and decamers at different temperatures. The less-resolved distribution $g_{ic}(r)$ drawn from the instantaneous dynamics is also plotted. Bonded monomers give a $\delta$-like contribution at $r=b=0.97$. For larger distances, three peaks at $r_1$, $r_2$ and $r_3$ are apparent. 
Fig. \ref{figConfoBond} shows the arrangements of the three bonded monomers resulting in the distances $r_1$, $r_2$ and $r_3$ for $M=3,10$. 
Notice that the adjacent bonds in each chain fragment shown in Fig. \ref{figConfoBond} are the RBs of the central monomer (see Fig.\ref{fig0}).
The first and highest peak of $g_{ic}^{IS}(r)$ at $r_1\approx1.12$ corresponds to folded polymers in which two non-consecutive monomers are at a distance $r_1=\sqrt[6]{2}\approx 1.12$, the position of the minimum of the Lennard-Jones potential. The central peak at $r_2\approx1.69$ corresponds to the partially-folded conformation with two consecutive bonds forming an angle  $\theta\approx122^\circ$. The third peak at $r_3\approx 2b \approx1.94$ is due to the linear arrangement of three monomers. The peaks of the
intrachain radial distribution function exhibit weak temperature dependence, more apparent in the peak at $r_1$. The latter is higher for decamers in that the first-neighbour shell is enriched in monomers belonging to the same chain of the tagged monomer. Reminding that our model exhibit finite repulsive forces, it is expected - and actually found - that the peaks at $r_2$ and $r_3$ are broader than the corresponding ones of hard-sphere monomers \cite{LasoJCP09}. Fig. \ref{fig1} shows that the broadening increases with the chain length. The increased dispersion is a manifestation of the fact that the number of chain configurations leading to two non-bonded monomers at $r$ distance increases with the chain length and approaches a three-dimensional featureless gaussian form \cite{DoiEdwards}. A similar effect is found for the end-to-end distance \cite{LasoJCP09}. Another effect contributing to broaden the peaks at $r_2$ and $r_3$ is discussed in Sec.\ref{bondbond}.  It is worth noting that the weak dependence of structure on temperature in the inherent structures has been also  noted in simple  (like binary atomic mixtures \cite{AndersenPRL88})  and complex (like water \cite{SastrySciortinoStarrStanley00} ) glass formers.

 \begin{figure}[t]
\begin{center}
\includegraphics[width=0.5\linewidth]{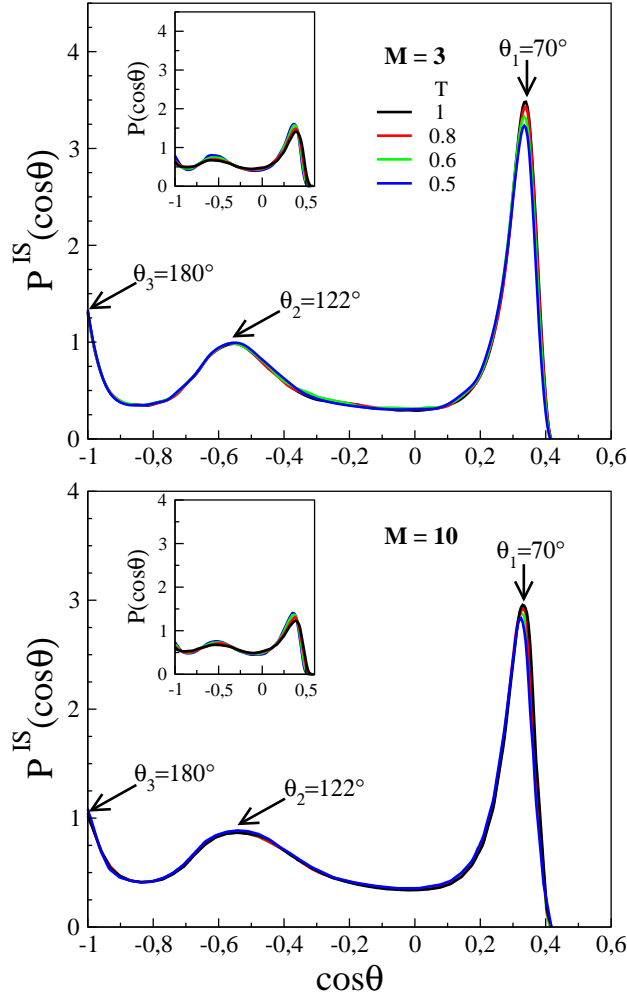} \\
\end{center}
\caption{ Inherent distribution of the angle between adjacent bonds $P^{IS}(\cos \theta)$ for trimers (top) and decamers (bottom) at the indicated temperatures (see Fig.\ref{fig0}). Inset: corresponding instantaneous distribution. 
$\theta_1\approx 70^\circ$, $\theta_2 \approx 122^\circ$ and $\theta_3\approx 180^\circ$ originate from the conformations presented in Fig.\ref{figConfoBond} a, b, and c respectively.}
 \label{figBond}
\end{figure}

\subsection{Radial bond orientational order}
\label{bondbond}
Let us consider the distribution $P(\cos \theta)$ of the angle between adjacent bonds in a chain, i.e. the RBs of a certain tagged monomer (see Fig. \ref{fig0}). In Fig.  \ref{figBond} the distribution $P^{IS}(\cos \theta)$ for trimers and decamers is shown together with $P(\cos \theta)$. Again, IS  configurations yield better resolution with respect to the instantaneous ones (Fig.\ref{figBond}, inset). The distribution vanishes above $\cos\theta \approx0.5$ due to monomer-monomer repulsion. 
The peaks of $P^{IS}(\cos \theta)$ at  $\theta_1\approx 70^\circ$, $\theta_2 \approx 122^\circ$ and $\theta_3\approx 180^\circ$ originate from the a), b) and c) conformations, respectively, evidenced by the intrachain part of the radial distribution function (Fig.\ref{figConfoBond}) \cite{noteBasch}. 

It must be stressed that the chains under study are {\it fully-flexible}, i.e. there is no torsional potential depending on the $\theta$ angle. Then,  the peaks of $P^{IS}(\cos \theta)$ have to be ascribed to local packing effects in FCSR.

Fig.  \ref{figBond} shows that the temperature effects on the distribution are minor and more visible in the narrower peak corresponding to $\theta_1\approx 70^\circ$. The widths of the three peaks increase a little with the chain length reflecting competition between the efficient local packing and the constraints due to the connectivity. The increase of the breadth of the peaks of $P^{IS}(\cos \theta)$ at $\theta_2 \approx 122^\circ$ and $\theta_3\approx 180^\circ$
with the molecular weight contributes to the analogous increase of the breadth of the peaks at  $r_2$ and $r_3$ of the intrachain inherent radial distribution function $g_{ic}^{IS}(r)$  (see Fig.\ref{fig1}). Differently, the peak of  $g_{ic}^{IS}(r)$ at $r_1$ is dominated by  other effects discussed in Sec. \ref{radial}.

\subsection{Steinhard's  bond orientational order}
\label{orderpar}

This Section presents the results on the FCS orientational order.  Sec. \ref{Generalities} defines the quantities of interest. Then,  the Section \ref{globalorderpar} illustrates the results concerning the order parameters averaged over the {\it whole} ensemble of monomers. 
It will be shown that FCS order is much lower than the one of the atomic liquids. To gain insight into this disordering effect,
the next two sections, \ref{radialorderpar} and \ref{tangentialorderpar}, concentrate on the order parameters averaged over {\it specific} fractions of interest. Sec. \ref{radialorderpar} considers the fractions of monomers with RBs forming the characteristic bond-bond angles $\theta \approx 70^\circ, 122^\circ, 180^\circ$ (see Fig.\ref{figConfoBond}). Sec. \ref{tangentialorderpar} considers the fractions of monomers with FCS having different number of SBs.

\subsubsection{Generalities}
\label{Generalities}

To gain insight into the FCS orientational order, we resort to global and local measures of the orientational order  $Q_{l,global}$ and $Q_{l,local}$, respectively \cite{Steinhardt83}. To this aim, one considers in a given coordinate system the polar and azimuthal angles $\theta({\bf r}_{ij})$ and $\phi({\bf r}_{ij})$ of the vector ${\bf r}_{ij}$ joining the $i$-th central monomer with the $j$-th one belonging to the neighbors within a preset cutoff distance $r^* = 1.2 \; \sigma^* \simeq 1.35$ \cite{Steinhardt83}. $r^*$ is a convenient definition of the FCS size \cite{sim}. The vector ${\bf r}_{ij}$ is usually referred to as a "bond" and has not to be confused with the {\it actual} chemical bonds of the polymeric chain!

To define a global measure of the order in the system, one calculates the quantity  \cite{Steinhardt83}:
\begin{equation} \label{Qbarlm_global}
	\bar{Q}_{lm}^{global}=\frac{1}{N_{b}}\sum_{i=1}^{N}\sum_{j=1}^{n_b(i)}Y_{lm}\left[\theta({\bf r}_{ij}),\phi({\bf r}_{ij})\right]
\end{equation}
where $n_b(i)$ is the number of bonds of $i$-th particle, $N$ is the total number of particles in the system, $Y_{lm}$ denotes a spherical harmonic and $N_b$ is the total number of bonds i.e:
\begin{equation} \label{N_b}
	N_b=\sum_{i=1}^{N} n_b(i) 
\end{equation}
 The global orientational order parameter $Q_{l,global}$ is defined by the rotationally invariant combination:
\begin{equation} \label{Ql_global}
 Q_{l,global}=\left [ \frac{4\pi}{(2l+1)} \sum_{m=-l}^{l} |\bar{Q}_{lm}^{global}|^2 \right ]^{1/2}
\end{equation}
It is interesting to consider the limit case of disordered systems where the bonds are not spatially correlated but they are distributed uniformly around a unit sphere. In that case one finds  \cite{RintoulTorquatoJCP96}:
\begin{equation} \label{Ql_global_uncorrelated}
Q_{l,global}^{uncor}= \frac{1}{\sqrt{N_b}} \pm \frac{1}{\sqrt{4 l + 2}} \frac{1}{\sqrt{N_b}}
\end{equation}
where the rightmost term on the right hand side of Eq. \ref{Ql_global_uncorrelated} is the expected width of the fluctuations.

\begin{figure}[t]
\begin{center}
\includegraphics[width=0.9\linewidth]{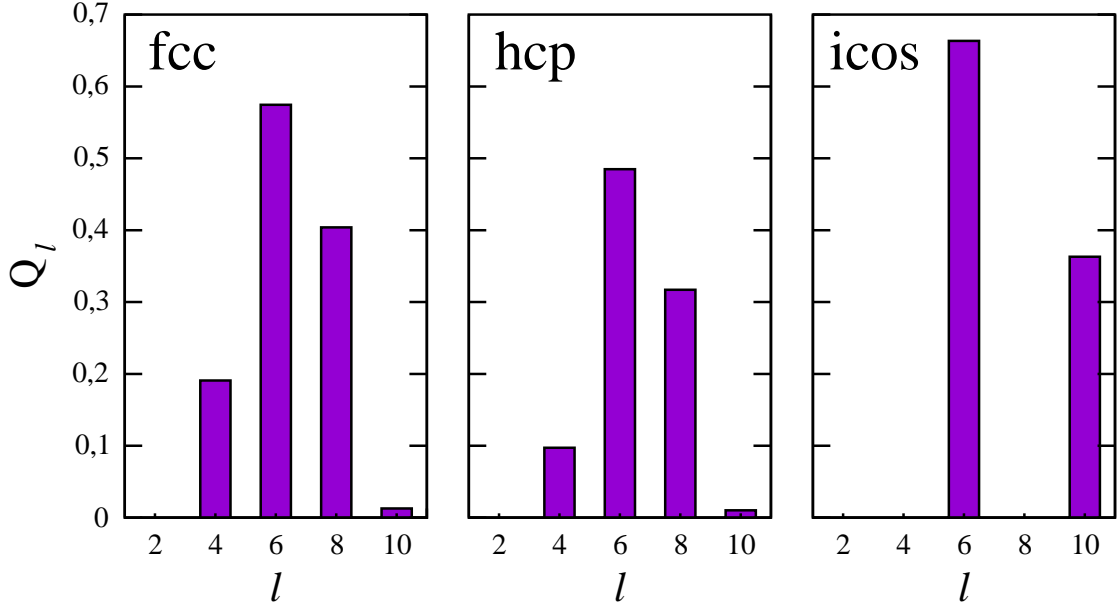} \\
\end{center}
\caption{Orientational order parameters $Q_l$ for systems with face-centered cubic (fcc), hexagonally close-packed (hcp) and icosahedral (icos) order.}
 \label{figQideal}
\end{figure}

A local measure of the orientational order  is obtained by considering the bonds between the $i$-th monomer and its $n_{b}(i)$ neighbours. To this aim, one calculates the quantity :
\begin{equation} \label{Qbarlm_local}
	\bar{Q}_{lm}^{local}( i )=\frac{1}{n_b(i)}\sum_ {j=1}^{n_b(i)}Y_{lm}\left[\theta({\bf r}_{ij}),\phi({\bf r}_{ij})\right]
\end{equation}
The local order parameters $Q_{l,local}$ is defined as \cite{Steinhardt83}:
\begin{equation} \label{Ql_local}
 Q_{l,local}=\frac{1}{N} \sum_{i=1}^{N}  \left [ \frac{4\pi}{(2l+1)} \sum_{m=-l}^{l} |\bar{Q}_{lm}^{local}( i )|^2 \right ]^{1/2}
\end{equation}
It has been noted that the choice $r^*\approx1.2 \; \sigma^*$, originally proposed in ref.\cite{Steinhardt83},  overestimates $Q_{l,local}$ with respect to other alternative neighborhood definitions  \cite{VoronoiMeckeJCP13}.

Twelve hard spheres may be put in simultaneous contact with a centre sphere in three different ways, two of them are characteristic of the face-centered cubic (fcc) and  hexagonally close-packed (hcp) crystals, whereas the third one (icos) is characteristic of the icosahedral arrangement and is unable to fill the space by replication \cite{FrankPRSL52}. 
 In  Fig.\ref{figQideal} the order parameters of ensembles of particles where the neighbors are arranged with  fcc, hcp and icos order are shown.  Non-zero values appear for $l\geqslant 4 $ in the fcc and hcp arrangements while in the icosahedral system non-zero values occur  for $l=6$ and $l=10$ only. 
 It must be pointed out that for systems where {\it all} the particles have the {\it same} neighborhood configuration the equality  $Q_{l,global}=Q_{l,local}= Q_{l}$ holds.  

\begin{figure}[t]
\begin{center}
\includegraphics[width=0.5 \linewidth]{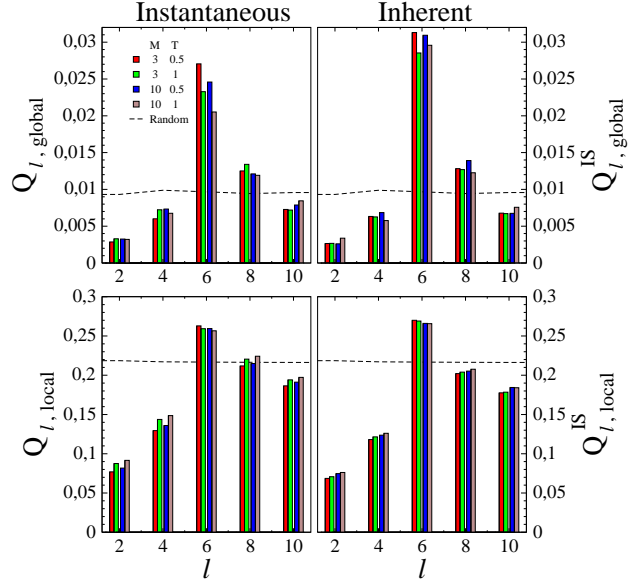} \\
\end{center}
\caption{Orientational order parameters $Q_{l,global}$ (top panel) and $Q_{l,local}$ (bottom panel) for instantaneous (left column) and inherent (right column) configurations of trimers and decamers at temperatures $T=0.5, 1$. The dashed lines connect the corresponding values $Q_{l,local}^{random}$, $Q_{l,global}^{random}$ for random bond configuration.}
 \label{figQ}
\end{figure}
\begin{figure}[b]
\begin{center}
\includegraphics[width=0.5\linewidth]{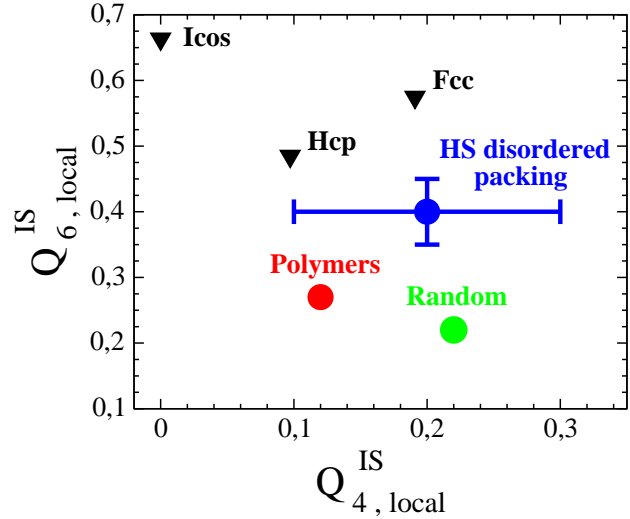} \\
\end{center}
\caption{$Q_{6,local}^{IS}$ vs. $Q_{4,local}^{IS}$ of the inherent configurations of decamers at $T=0.5$ (red dots).  $Q_{4,local}^{IS} = 0.12$, $Q_{6,local}^{IS} = 0.27$. The pair $Q_{4,local}^{random} \simeq  Q_{6,local}^{random} = 0.22$ of random bond configurations is also plotted (green dots). The blue dots and related error bars denote the regions spanned by the most recurrent values of the $\{Q_{4,local}^{HS} , Q_{6,local}^{HS} \}$ pairs in hard-spheres packings
($0.10 \le Q_{4,local}^{HS} \le 0.30$ and $0.35 \le Q_{6,local}^{HS} \le 0.45$) \cite{AsteJPCM05,AstePRE05}. The triangles denote the points for fcc ($Q_{4}^{fcc}=0.191$, $Q_{6}^{fcc}=0.574$), hcp ($Q_{4}^{hcp}=0.097$, $Q_{6}^{hcp}=0.485$) and icos  ($Q_{4}^{icos}=0, Q_{6}^{icos}=0.663$) clusters (see Fig.\ref{figQideal}). 
} 
 \label{figQ6Q4}
\end{figure}

\subsubsection{Global and local order in the polymer melt}
\label{globalorderpar}

Fig.\ref{figQ} shows the global (top panels) and the FCS local (bottom panels) order parameters of a melt of trimers and decamers  at $T=0.5,1$ due to both their instantaneous (left panels) and inherent (right panels) configurations. The values are averaged over {\it all} the monomers.
For comparison, the order parameters, $Q_{l,global}^{random}$ and $Q_{l,local}^{random}$, of a random neighborhood configuration around a centre particle are also plotted. The random configuration has been built by considering  the $n_b(i)$ neighbors of the $i$-th particle and replacing  the actual bond orientation  $\{ \theta({\bf r}_{ij}),\phi({\bf r}_{ij})\}$ of the $j$-th neighbor with a fictitious random one. This yields $Q_{l,global}^{random} \simeq 1.0 \cdot 10^{-2}$ and $Q_{l,local}^{random} \simeq 2.2 \cdot 10^{-1}$.
According to eq.\ref{Ql_global_uncorrelated} and our sample size, one expects $Q_{l,global}^{random} \simeq Q_{l,global}^{uncor} \sim 1/ \sqrt{12 \cdot 500} \sim 1.3 \cdot 10^{-2}$ and $Q_{l,local}^{random} \simeq Q_{l,local}^{uncor} \sim 1/ \sqrt{12} \sim 2.9 \cdot 10^{-1}$. 
It is seen in Fig.\ref{figQ} that the global order of the sample is small, but not completely negligible due to the finite sample size.  It is comparable to the one observed  for an atomic liquid with Lennard-Jones potential at $T=0.719$  \cite{Steinhardt83}. It is apparent that, on cooling, the temperature dependence is very weak, thus differing from the behaviour of an atomic liquid \cite{Steinhardt83}. The global order is also found to be negligibly dependent on the chain length.
Fig.\ref{figQ} shows that the instantaneous and the mechanically-equilibrated inherent configurations yield very similar order parameters, thus suggesting minor role by the vibrational motion.

Fig.\ref{figQ} evidences that the FCS local order is quite low with respect to the ideal cases (Fig.\ref{figQideal}) even at the lowest temperature under study and the order parameters are very close to the random values $Q_{l,local}^{random}$, especially for $l=6,8,10$. This aspect is better seen by plotting the $(Q_{4,local}^{IS}$, $Q_{6,local}^{IS})$ pairs in Fig.\ref{figQ6Q4}. This evidences that the connectivity tends to reduce the local order which not only departs markedly by the icos, hcp and fcc ordering, but it also exhibits significant deviations from the HS disordered packing. 

Insight into the disordering effect due to the connectivity is provided by the next two sections. Owing to the small dependence of the local order on both the temperature and the chain length, henceforth, only the decamer melt at $T=0.5$ will be considered.

\begin{figure}[t]
\begin{center}
\includegraphics[width=0.5\linewidth]{fig8.eps} \\
\end{center}
\caption{Local order parameters of the fractions of tagged monomers with RBs forming the angle $\theta  =   70^\circ, 122^\circ, 180^\circ$ within $5 \%$ in the melt of decamers at $T=0.5$ (see Figs.\ref{fig0},\ref{figBond}).  The results are compared with the average values of all the monomers. The virtually $\ell$-independent order parameters of the totally random FCS are also indicated.
} 
 \label{radialbond}
\end{figure}

\begin{figure}[t]
\begin{center}
\includegraphics[width=0.5 \linewidth]{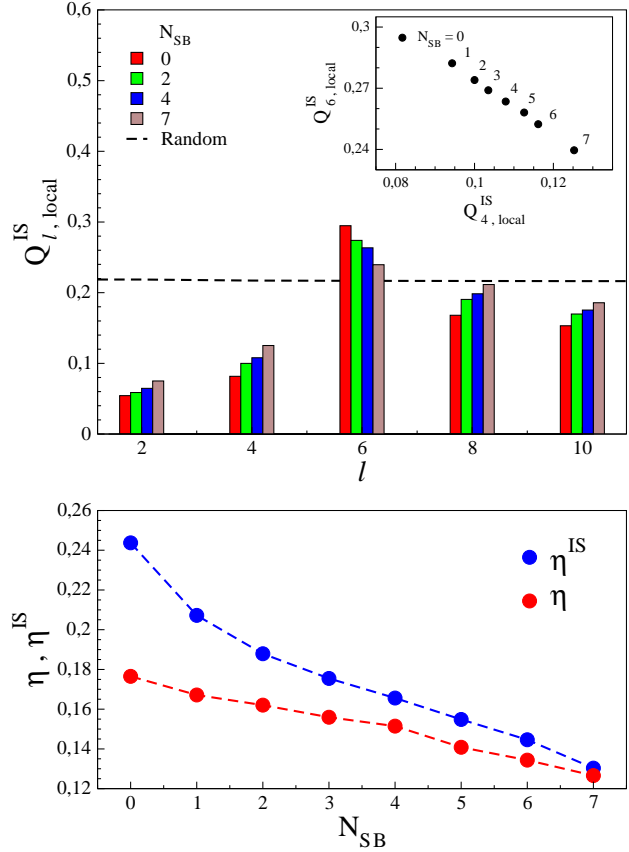} \\
\end{center}
\caption{Top: local order parameters of the fractions of tagged monomers with different number $N_{SB}$ of shell bonds in their FCSs  (see Figs.\ref{fig0},\ref{figBond}).  The results refer to the melt of decamers at $T=0.5$. The virtually $\ell$-independent order parameters of the totally random FCS are also indicated. Inset: $Q_{6, \, local}^{IS}$-$Q_{4, \, local}^{IS}$ plot of the different fractions (compare with the bulk behavior in Fig.\ref{figQ6Q4}). Bottom: instantaneous $\eta$ and inherent $\eta^{IS}$ measures (Eq.\ref{measure}) of the increasing FCS  ordering by removing shell bonds. 
} 
 \label{shell}
\end{figure}

\subsubsection{Influence of the radial bonds on the local order}
\label{radialorderpar}
The previous section refers to the order parameters averaged over {\it all} the monomers. We now consider 
the order parameters distinctly averaged over  the three fractions of monomers with RBs forming the characteristic bond-bond angles $\theta \approx 70^\circ, 122^\circ, 180^\circ$ (see Fig.\ref{figConfoBond}).
Fig. \ref{radialbond} shows the FCS order parameters of their IS configurations for a melt of decamers at $T=0.5$. It is shown that the FCS order of the fractions little differ from the ones of the bulk, in particular if $\ell \neq 6$. This is understood in that each RBs involve only one of the about twelve monomers that, on average, form one FCS. Then, they are unable to compete against the cooperative effort setting the FCS arrangements. Instead, as discussed in Sec. \ref{bondbond}, the RB orientational order is set by the FCS orientational order.

It is seen that the differences with respect to the random configuration increase with the bond-bond angle. This effect may be rationalized in that, on decreasing the bond-bond angle, the increasing localization of the two RBs reduces their perturbing effects. The conclusion is consistent with the finding that the fraction of  the monomers located in the chain end, which have only {\it one} RB, exhibits FCS order parameters virtually identical to the ones of the fraction  with {\it two} RBs forming the bond-bond angle $\theta \simeq   70^\circ$ (data not shown in Fig.\ref{radialbond} for clarity reasons).

\begin{figure}[t]
\begin{center}
\includegraphics[width=0.35\linewidth]{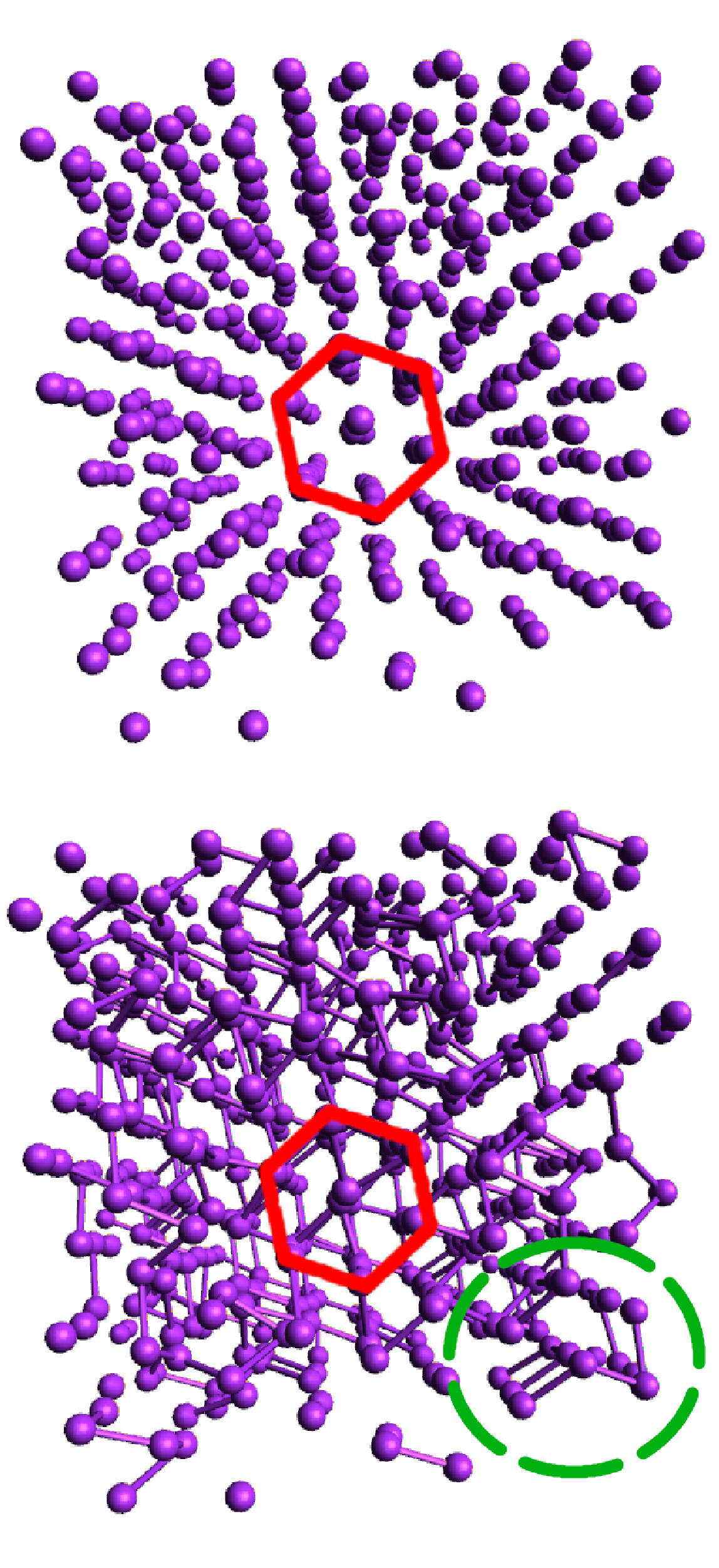} \\
\end{center}
\caption{View along the $c$ (chain) axis of the inherent structure of the decamer crystal at $T=0.75$, $\rho = 1.086$ by hiding (top) or not (bottom) the bonds between the monomers. For clarity reasons, the monomer 
size is smaller than the actual one. The top panel emphasizes the hexagonal order of the columns of piled monomers, whereas the bottom panel evidences the transverse bonds to the $c$ axis. The latter form localized defect, known as kinks \cite{Gedde95,PolymCrystJChemEngData02}, which put monomers of the {\it same} chain in  {\it different} columns. { The highlighted region close to the right lower corner of the bottom panel shows a four-monomer portion of a decamer traversing the top of three columnar fragments. The chain portion has one bond-bond angle $\sim 120^\circ$ (the leftmost) and the other one  $\sim 70^\circ$ (the rightmost).}
Notice that, due to the periodic boundary conditions of the simulation box,  some monomers look like as  isolated and some chains fragmented.
}
 \label{cryst}
\end{figure}

\subsubsection{Influence of the shell bonds on the local order}
\label{tangentialorderpar}

Fig.\ref{shell} (top panel) shows the dependence of the local order on the number of shell bonds $N_{SB}$ in FCS of the tagged monomer. It is seen that, on decreasing $N_{SB}$, the order parameters of the IS configurations deviate more and more from the ones of a totally disordered FCS, i.e. the local order increases.  By comparison with Fig.\ref{radialbond}, it is seen that  the effect of changing the SB number in the FCS is stronger than changing the RBs angular configurations. In particular, changes are quite apparent for {\it all} the order parameters, whereas RBs mostly affect $Q_{6,local}^{IS}$, see Fig.\ref{radialbond}.

The inset of the top panel of Fig.\ref{shell} plots the $Q_{6, \, local}^{IS}$-$Q_{4, \, local}^{IS}$ pairs of the main panel. If compared with the same plot of the bulk system, Fig.\ref{figQ6Q4},  it suggests that removing SBs from FCS increases the FCS orientational order  with features of the HCP and ICOS atomic ordering more than the FCC one. 

To quantify the ordering we define the measure:
\begin{equation}
\eta = \frac{1}{5} \sum_{\ell} \left ( \frac{Q_{\ell, \, local} - Q_{\ell,\, local}^{random} }{Q_{\ell,\, local}^{random}} \right )^2  \label{measure}
\end{equation}
If the above quantity is evaluated in terms of the IS configurations is referred to as $\eta^{IS}$. Fig.\ref{shell} (bottom panel) shows the decrease of both $\eta^{IS}$ and $\eta^{IS}$ by increasing the SB number in FCS. In particular, it is seen that $\eta^{IS} \simeq \eta$ for FCS with high $N_{SB}$.  In fact, the high number of SBs stiffen FCS, making less pronounced the differences between the instantaneous and the IS configurations.

\begin{center}
\begin{table}[t]
\caption{Local and global order  parameters ($l=4,6$) of the decamer melt and crystalline states at $T=0.75, \rho = 1.086$. The HPC entry refers to  an ideal ensemble of hexagonally-packed columns of piled monomers not connected by bonds.
\label{parametri}}
\begin{ruledtabular}
\begin{tabular}{ccccc}
\vspace{0.05mm}\\ 
 & $Q_{4,local}^{IS}$ & $Q_{6,local}^{IS}$ & $Q_{4,global}^{IS}$ & $Q_{6,global}^{IS}$   \\ \\
 \hline
\vspace{0.05mm}\\
Melt   & $0.11$   & $0.26$   & $0.004$ & $0.039$    \\ \\
Crystal   & $0.083$   & $0.31$   & $0.031$ & $0.25$  \\ \\
HPC   & $0.031$   & $0.27$   & $0.031$   & $0.27$
\end{tabular}
\end{ruledtabular}
\end{table}
\end{center}

 \begin{figure}[b]
\begin{center}
\includegraphics[width=0.5\linewidth]{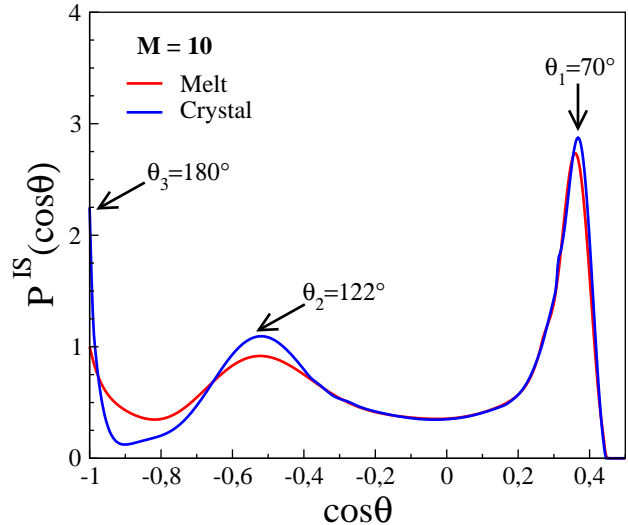} \\
\end{center}
\caption{Comparison between the inherent distribution of the angle between adjacent bonds $P^{IS}(\cos \theta)$ of decamers in the melt and the  crystal both at $T=0.75, \rho = 1.086$.}
 \label{pcostetacrysta}
\end{figure}

\subsection{Influence of the global order on the local order}
\label{global}
This Section investigates how the global order affects the local order. To this aim,  the decamer melt at temperature $T=0.75$ and density $\rho = 1.086$ is compared with the crystalline state being occasionally formed during a few of the equilibration runs at the same temperature and density.
 Fig. \ref{cryst} (top) shows the monomer arrangements of the decamer crystal. Monomers are packed in linear columns (defining the crystal $c$ axis) which are hexagonally packed in the $ab$ orthogonal plane. This kind of arrangement of the flexible and the semi-flexible chains is seen in experiments  (for reviews see \cite{PolymCrystJChemEngData02,AllegraPolymCrystHex04,CorradiniPolCryst05}) and simulations of bulk assemblies  \cite{PolCrystTakeuchi98,OctamerCrysJCP12,MilnerJCP11}, as well as isolated single molecules \cite{LariniCrystJPCM05,LariniCrystJCP05}. 

\begin{figure}[t]
\begin{center}
\includegraphics[width=0.5\linewidth]{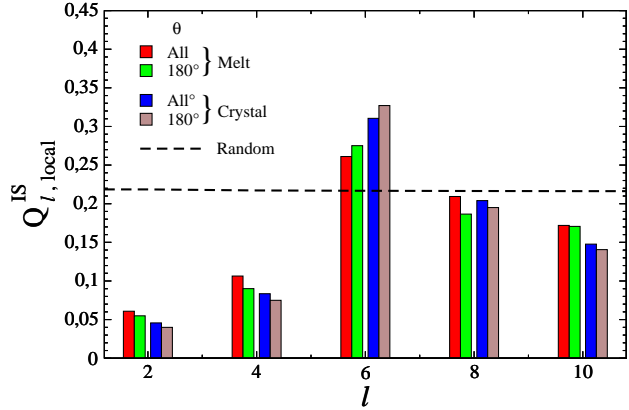} \\
\end{center}
\caption{Local order parameters of the fraction of tagged monomers with RBs forming the angle $\theta  =   180^\circ$ within $5 \%$ in the melt and the crystalline states of decamers  both at $T=0.75, \rho = 1.086$ (see Fig.\ref{fig0}).  
The results are compared with the average values of all the monomers. The virtually $\ell$-independent order parameters of the totally random FCS are also indicated.
} 
 \label{radialbondCrys}
\end{figure}

\begin{figure}[t]
\begin{center}
\includegraphics[width=0.5 \linewidth]{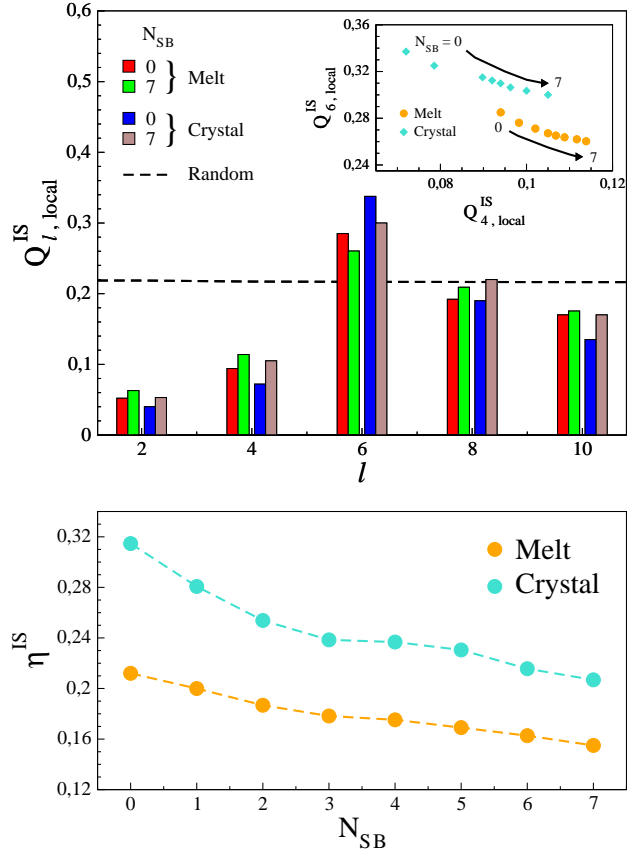} \\
\end{center}
\caption{Top: local order parameters of the fractions of tagged monomers with minimum  ($N_{SB} = 0$) or maximum  ($N_{SB} = 7$) number of  SBs in  FCS (see Fig.\ref{fig0}).  The results refer to the melt and the crystalline states of decamers  both at $T=0.75, \rho = 1.086$. The virtually $\ell$-independent order parameters of the totally random FCS are also indicated. Inset: $Q_{6, \, local}^{IS}$-$Q_{4, \, local}^{IS}$ plot of the fractions with $0 \le N_{SB} \le 7$. Bottom:    $\eta^{IS}$ measure (Eq.\ref{measure}) of the decreasing FCS  ordering in fractions with increasing number of SBs.  The effect is more marked in the crystalline state. 
} 
 \label{shellCrys}
\end{figure}

 To make more quantitative the visual impression of Fig. \ref{cryst} (top), Table \ref{parametri} lists the order parameters with $l=4,6$ of the crystalline state and the melt  together with the ones of an ideal ensemble of hexagonally-packed columns of piled monomers not connected by bonds (HPC). It is seen that  the global order parameters $Q_{4,global}^{IS}$, $Q_{6,global}^{IS}$ of the crystalline state are larger of a factor of about $8$ and $6$ respectively with respect to the melt state and compare rather well with the HPC model. The HPC estimate is satisfying for $Q_{6,local}^{IS}$ but poor for $Q_{4,local}^{IS}$. All in all, given the high ideality of the HPC model, which does not take into account the disturbing effect of the connectivity, the overall agreement is encouraging. Notice that perfectly ordered states have $Q_{l,global} = Q_{l,local}$. This equality roughly holds for $l=6$ of the crystal state but fails for $l=4$ pointing to the presence of some degree of disorder. 
  
 One  feature of  the crystal phase is the presence of in-chain point-like (zero dimensional) defects. In fact, Fig. \ref{cryst} (bottom) shows that some of the bonds are not aligned to the  column $c$ axis so that monomers of the {\it same} chain belong to  {\it different} monomer columns. This defect is well known in polymer crystals and usually referred to as a "kink" \cite{Gedde95,PolymCrystJChemEngData02}.  We ascribe the presence of defects to the fully-flexible character of the model polymer under study. Notably, kinks  are not evidenced in simulations concerning the crystalline state of linear molecules with higher stiffness  due to torsional and bond angle-bending potentials  \cite{PolCrystTakeuchi98,OctamerCrysJCP12,MilnerJCP11}.
 
More insight about the chain conformations  is gained by comparing in Fig.\ref{pcostetacrysta} the crystal and the melt with respect to  their inherent distributions of the angle between adjacent bonds. We remind that adjacent bonds are the RBs of the central monomer from which  both of them depart (see Fig.\ref{fig0}) and then are involved in the FCS local order.  
Fig.\ref{pcostetacrysta} shows that the crystallization alters the distribution $P^{IS}(\cos \theta)$ virtually only in the region $120^\circ \lesssim \theta \lesssim 180^\circ$ by increasing the fractions of bond-bond angles with $\theta \sim 120^\circ$ and  $180^\circ$ with respect to the melt. Both $\theta$ values are compatible with hexagonally-packed columns of piled monomers and the presence of bonds not aligned to the column axis, 
{ e.g. see the highlighted region close to the right lower corner of Fig. \ref{cryst} (bottom) where three columnar fragments are traversed by one four-monomer portion of a decamer with one bond-bond angle $\sim 120^\circ$}. 
Misaligned bonds with $ 70^\circ \lesssim \theta \lesssim 110^\circ$ also contribute to $P^{IS}(\cos \theta)$. This fraction (about $60\%$) is nearly unaffected by the crystallization. Even if more work, beyond the purpose of the present paper, is needed to clarify the issue, two tentative explanations are offered. First, one notice that  $25 \%$ of the bond-bond angles involve chain ends which are surrounded by disordered regions loosely coupled to the crystalline regions. Second, one guesses 
that folded arrangements with $ \theta \sim 70^\circ$  (Fig.\ref{figConfoBond}a) are accommodated with limited strain in the ordered structure, perhaps by a local distortion of the hexagonal column packing, { as suggested by the small shift to smaller $\theta$-values of the peak at $ \theta \sim 70^\circ$ of $P^{IS}(\cos \theta)$ of the crystal and the highlighted region close to the right lower corner of Fig. \ref{cryst} (bottom) where three columnar fragments are traversed by one four-monomer portion of a decamer with one bond-bond angle $\sim 70^\circ$.}

Does the global order change the local order and the influence on it  by the RBs and SBs ? These issues are addressed below.

Fig. \ref{radialbondCrys} compares  the local order parameter of the melt and the crystalline state of the decamer. One considers both the average values and the ones pertaining to  the fraction of monomers with collinear RBs ($\theta =180^\circ$).  It is seen that the FCS order  of the crystal deviates from the one of the random FCS more than the melt, i.e the increased global order improves the local order too. A shared feature of the crystal and the melt  is the fact the FCS order of the fraction with collinear RBs little differs from the ones of the bulk. This suggests that, since RBs of the crystalline state involve - as in the melt - only  a few FCS monomers, they are unable to affect the FCS arrangements significantly. Instead, as Fig.\ref{pcostetacrysta} shows, the RBs arrangements are, at least in part, set by the FCS arrangements enforced by the global order.

Fig. \ref{shellCrys} compares the local order  of the melt and the crystal for different numbers of SBs in FCS (see Fig.\ref{fig0}).  As in the melt, the presence of SBs in FCS decreases the local order of the crystal. The inspection of the order parameters shows that the effect is fairly larger in the crystal. This is emphasized in the inset of  the top panel of Fig. \ref{shellCrys} where one sees that increasing the SB number in FCS of monomers in the crystalline sample results in larger shifts of the representative point in the $Q_{6, \, local}^{IS}$-$Q_{4, \, local}^{IS}$ plane. Another way to appreciate the larger influence of the SBs in the local order is to consider the measure $\eta^{IS}$ (Eq.\ref{measure}). This is done in the bottom panel of Fig. \ref{shellCrys}. It is seen that changing the number of SBs from the minimum to the maximum values changes $\eta^{IS}$ of about $26 \%$ in the melt and about $36 \%$ in the crystalline phase.

\section{Conclusions}
\label{conclusions}

The present paper investigates the competition between the connectivity and  the FCS local order by a MD study of model fully-flexible chain molecules ($M=3, 10$).  States  with both missing  (melts) and high (crystal) global order are considered and compared. 
The crystalline state is characterized as an hexagonally-packed assembly of columns of piled monomers.  In-chain point-like (zero dimensional) kink defects are present in the ordered structure and force monomers of the {\it same} chain to belong to  {\it different} monomer columns.  The presence of defects is ascribed to the fully-flexible character of the model polymer under study. 
The changes in FCS ordering  have been analysed in terms of  Steinhardt's orientation order parameters. The latter, if averaged over all the monomers, do not change appreciably with both the chain length and the temperature. The FCS ordering of the connected systems is significantly lower than the crystalline, quasi-crystalline and disordered arrangements of atomic systems. 

Insight into the decrease of the FCS order due to the connectivity is reached by considering the perturbing effects of the chains bonds, as divided in  two families, RBs and SBs.
SBs have deep influence on the FCS ordering, especially in the crystalline state. Indications suggest that their removal favours FCS ordering with features of the HCP and ICOS atomic ordering more than the FCC one. On the other hand, FCS ordering weakly depends on both the number and the mutual orientation of RBs. Even if the oligomer and the polymer chains are fully flexible, the distribution of  angles between adjacent radial bonds exhibits sharp contributions at  the characteristic angles $\approx 70^\circ, 122^\circ, 180^\circ$.  They are enhanced by the global order of the crystal if $\theta \approx  122^\circ, 180^\circ$, { whereas the distribution is nearly unaffected by the crystallization if  $ 70^\circ \lesssim \theta \lesssim 110^\circ$}. It is suggested that, since RBs involve only a few of the monomers that form one FCS, they are unable to compete against the cooperative effort setting the FCS arrangements.


\begin{thebibliography}{55}
\bibitem[{not()}]{noteBasch}
\bibinfo{howpublished}{Note that the alternative distribution $P'(\theta) =
  P(\cos \theta) \sin \theta$ misses the peak due to the linear configuration
  ($\theta \approx 180^\circ$) owing to the weight factor $\sin \theta$
  \cite{BaschnagelEPJE11}.}
\end{thebibliography}
\end{document}